\documentclass[longbibliography,
 reprint,
 amsmath,amssymb,
 aps,
]{revtex4-1}

\usepackage{color}
\usepackage{float}
\usepackage[normalem]{ulem}
\usepackage{graphicx}

\newcommand{\glv}{\gamma_{lv}}
\newcommand{\h}{h}
\renewcommand{\rm}{r_m}
\newcommand{\rs}{R_s}

\newcommand{\tsound}{t_{\mathrm{sound}}}
\newcommand{\Rf}{R_f}
\newcommand{\beq}{\begin{equation}}
\newcommand{\eeq}{\end{equation}}
\newcommand{\upd}{\mathrm{d}}

\newcommand{\lambdabarR}{\bar{\lambda}}
\newcommand{\Ksub}{K_{\mathrm{sub}}}
\newcommand{\Kdyn}{K_{\mathrm{dyn}}}

\definecolor{marker12}{RGB}{39,151,235}
\definecolor{marker22}{RGB}{101,205,110}

\begin{document}

\preprint{APS/123-QED}

\title{Dynamics of wrinkling in ultrathin elastic sheets}

\author{Finn Box$^1$}
\author{Doireann O'Kiely$^1$}%
\author{Maxime Inizan$^{1}$}
 \author{Ousmane Kodio$^1$}
  \author{Alfonso A. Castrej\'{o}n-Pita$^2$}
  \author{Dominic Vella$^1$}
\affiliation{$^1$Mathematical Institute, University of Oxford, Oxford OX2 6GG, United Kingdom}
\affiliation{$^2$Department of Engineering Science, University of Oxford, Oxford OX1 3PJ, United Kingdom}

\begin{abstract}
The wrinkling of thin elastic objects provides a means of generating regular patterning at small scales in applications ranging from photovoltaics to microfluidic devices. Static wrinkle patterns are known to be governed by an energetic balance between the object's bending stiffness and an effective substrate stiffness, which may originate from a true substrate stiffness or from tension and curvature along the wrinkles. Here we investigate dynamic wrinkling, induced by  the impact of a solid sphere onto an ultra-thin polymer sheet floating on water. The vertical deflection of the sheet's centre induced by impact draws material radially inwards,  resulting in an azimuthal compression that is relieved by the wrinkling of the entire sheet. We show that this wrinkling is truly dynamic, exhibiting features that are qualitatively different to those seen in quasi-static wrinkling experiments. Moreover, we show that the wrinkles coarsen dynamically because of the inhibiting effect of the fluid inertia. This dynamic coarsening can be understood heuristically as the result of a  dynamic stiffness, which  dominates the static stiffnesses reported thus far, and allows new controls of wrinkle wavelength. 
\end{abstract}



\maketitle


Wrinkling provides a means of reconfiguring slender structures \cite{Pellegrino,Reis2015}, and offers new opportunities for characterizing materials through thin sheet metrology \cite{Stafford2004,Huang2007,Ripp2018}. Control of mechanical properties also permits wrinkle orientation and geometry to be tailored, providing a simple and robust patterning method that can produce periodic structures with regular spacing that ranges from hundreds of nanometres to millimetres  \cite{Schweikart2009,Bayley2014}.  This has proved particularly versatile at small scales as an alternative to lithographic techniques: wrinkle formation on soft surfaces has been used to fabricate close-packed nanofluidic channels \cite{Chung2008}, surfaces with anisotropic wetting properties \cite{Chung2007}, ordered arrays of self-assembled colloidal particles \cite{Lu2007} and optical phase gratings \cite{Bowden1999,Harrison2004}.

The formation of wrinkles is induced by compression, with the critical compression and the emergent wavelength of wrinkles depending on a balance between the resistance to bending of the sheet and a stiffness that resists out-of-plane deformation (which may come from a substrate, or tension and curvature along the wrinkles \cite{Cerda2003,Huang2007,Davidovitch2011,Paulsen2016} as well as geometrical confinement \cite{Davidovitch2019,SteinMontalvo2019}). In applications, the wrinkle wavelength is often controlled by changing the sheet thickness (e.g.~through oxidation of thin silica layers \cite{Bowden1999,Harrison2004,Chung2007,Chung2008}). However, this wavelength is then set once and for all, and does not change significantly from its value at onset \cite{Cai2011}; while the amplitude of wrinkles \emph{can} be varied by further compression, applications are somewhat limited by this inability to generate different wavelength structures in the same system.

\begin{figure}
\begin{center}
\centering
\includegraphics[width=0.9\linewidth]{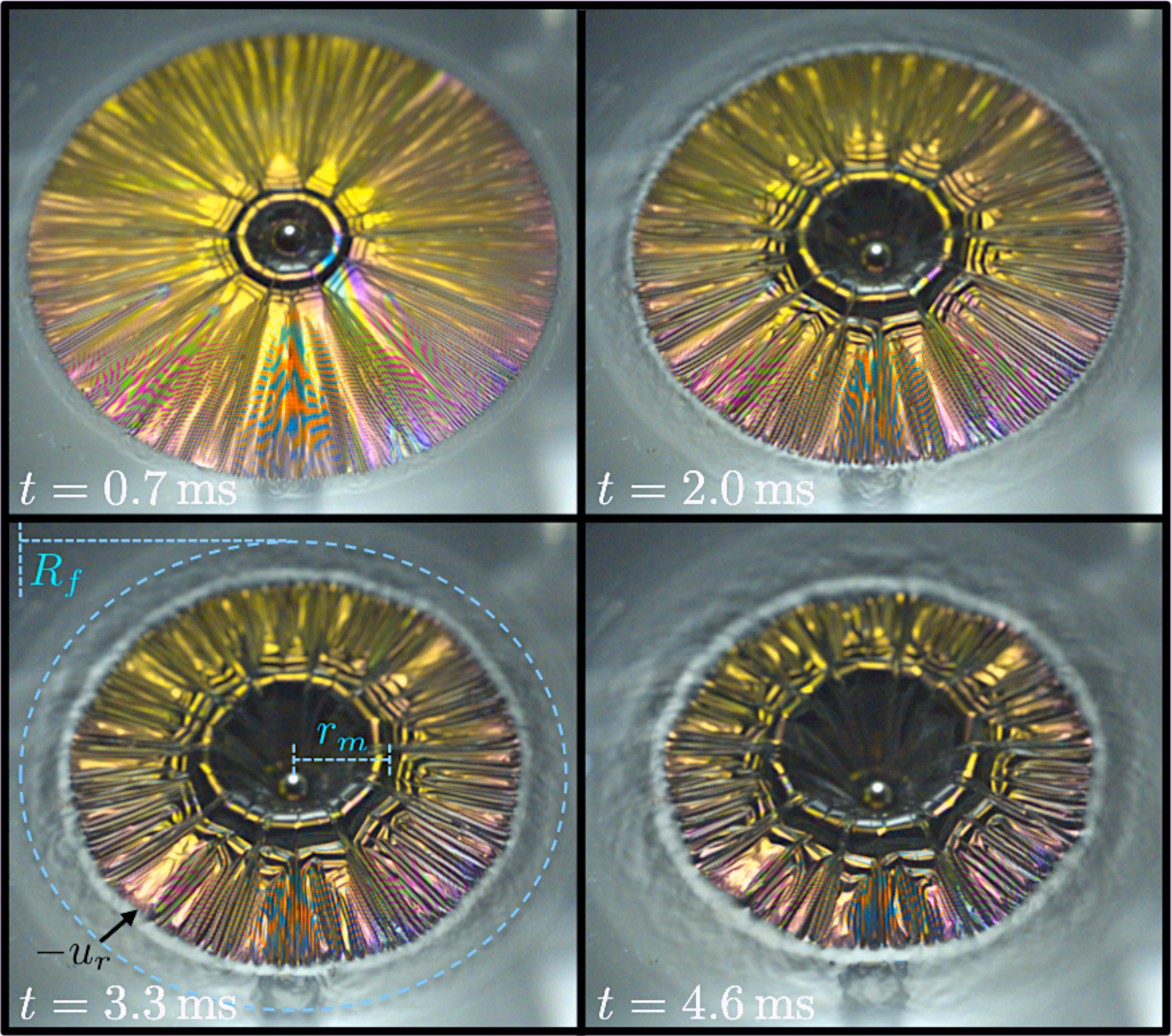}
\caption{A steel sphere (radius $\rs = 1.25$\,mm) impacts a Polystyrene (PS) sheet (thickness $\h=350\mathrm{~nm}$, radius $\Rf = 17.15\mathrm{~mm}$, floating at a water--air interface) at speed $V=0.72$\,m\,s$^{-1}$.  Impact draws the outer edge of the sheet inwards by a distance $u_r(\Rf,t)$ compressing the sheet even in regions where it remains flat (beyond the propagating transverse wave at $r=\rm(t)$). The number of radial wrinkles in the sheet decreases with time, i.e., with increasing impact depth $Vt$, in contrast to the static indentation of a floating sheet \cite{Holmes2010,Paulsen2016}.  
} 
\label{fig:colour}
\end{center}
\end{figure}

\begin{figure*}
\begin{center}
\includegraphics[width=\linewidth]{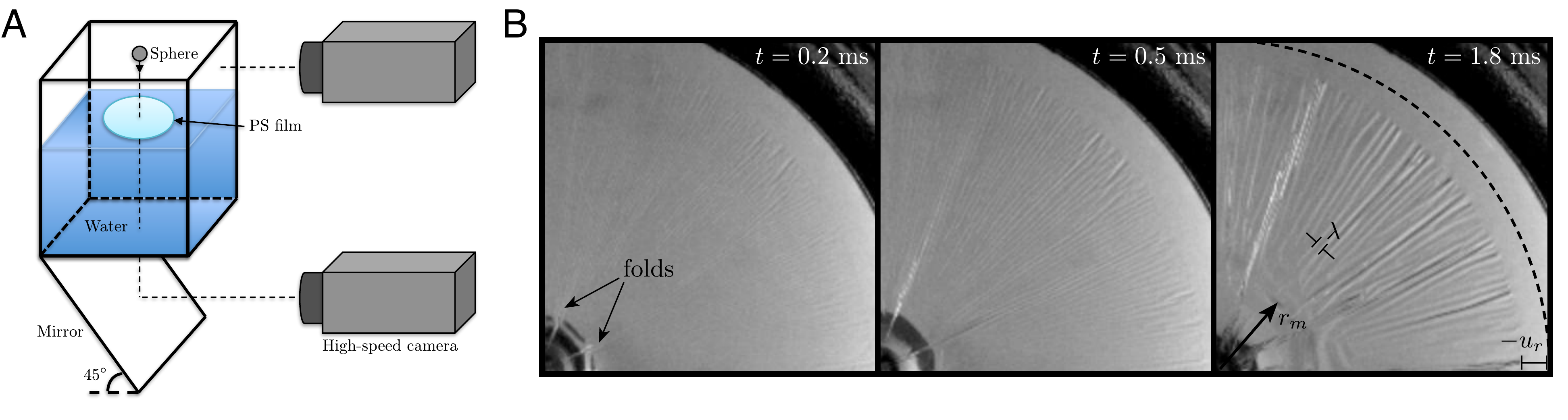} 
{\caption{(a) Schematic of the experimental setup used to drop steel spheres onto floating Polystyrene (PS) sheets. (b) Time series of one quadrant of a sheet ($\h=450$\,nm, $\Rf = 17.40$\,mm) impacted by a sphere (radius $\rs = 1.75$\,mm, $V=1.17$\,ms$^{-1}$) and imaged from below. Note that wrinkles coarsen during the experiment, trebling their mean wavelength in 1.5\,ms. Also highlighted in the images are the onset of folds, the propagation of a transverse wave (visible at radius $r_m$) and the retraction of the sheet edge (whose initial position   is shown as the dashed circle). }
\label{fig:schematic} }
\end{center}
\end{figure*}

A simple experiment that reveals some of the complexity of wrinkling is the indentation of an ultra-thin elastic sheet floating at a liquid--air interface \cite{Holmes2010,Vella2015,Paulsen2016,Vella2018,Ripp2018}. Indentation draws material radially inwards, in the process creating compressive stresses in the azimuthal direction. At a critical indentation, this compression overcomes the base capillary tension in the sheet \cite{Vella2015}, and radial wrinkles form (i.e.~wrinkles whose peaks and valleys lie along lines of increasing radius). Beyond this wrinkling threshold, the sheet rapidly wrinkles everywhere, but the pattern of wrinkles varies in different regions of the sheet: in a curved central portion, the wrinkle wavelength is controlled by curvature, while beyond this region, the sheet is flat and the hydrostatic pressure in the liquid instead controls the wrinkle wavelength.  In contrast to the case of uni-axial compression discussed above (where the wavelength does not vary with increased confinement), the wrinkles in the central curved portion of a poked sheet become more refined as the confinement is increased by further indentation~\cite{Paulsen2016}; ultimately these wrinkles form deep folds \cite{Holmes2010}. In the flat portion, however, the wrinkle wavelength remains constant as indentation progresses. This wavelength, $\lambda_0$, is set by the balance between the sheet's bending stiffness, $B$, and a substrate stiffness $\Ksub$, which gives~\cite{Cerda2003, Paulsen2016}:
\beq \label{eq:gravity}
\lambda_0=2\pi\left(\frac{B}{\Ksub}\right)^{1/4},
\eeq where the substrate stiffness appropriate to a liquid bath is  the specific weight of the liquid, i.e.~$\Ksub=\rho g$ with $\rho$ the liquid density and $g$ the acceleration due to gravity. In this article, we investigate how this static picture changes when indentation is performed dynamically, via impact.

Dynamic buckling instabilities have been investigated as a route for inducing pattern formation in rigid objects \cite{Vandeparre2010,Box2013}, and also as a route to understanding plastic crumpling in impacts \cite{Karagiozova2008} and  brittle fragmentation \cite{Gladden2005,Vermorel2007,Vermorel2009}. Impact on an elastic sheet has been studied both for a sheet in free-fall~\cite{Vermorel2009} and a sheet floating on the surface of water \cite{Duchemin2014,Vandenberghe2016}. In both cases, a longitudinal tensile wave propagates outwards from the point of impact at the speed of sound, stretching the sheet, followed by a transverse wave that propagates through the stretched domain. A coupling between these two waves leads to an azimuthal compression which gives rise to wrinkling, as expected from the static case.  In the flat region outside the transverse wave, the wrinkle wavelength is fixed, and is explained using the dynamic beam equation, with uniform imposed strain leading to a constant compressive force. Here, we use a similar experimental setup, but are able to create wrinkles that evolve dynamically during the course of the experiment, departing from observations in both static indentation experiments and previous dynamic impact experiments.

We investigate dynamic impact on ultra-thin polymer sheets subject to an applied background stress (provided by surface tension), see Fig.~\ref{fig:colour} and Movie S1. The experimental setup is illustrated in Fig.~\ref{fig:schematic}a.  Polystyrene (PS) sheets of thickness $150\mathrm{~nm}\leq \h\leq 530\mathrm{~nm}$, Young'sx modulus $E=3.46\mathrm{~GPa}$ and Poisson's ratio $\nu=0.33$ were created by spin coating a PS-in-toluene solution onto glass slides \cite{Huang2007}.  The resulting sheets were cut to have radii $5.5\mathrm{~mm}\leq\Rf\leq22.7\mathrm{~mm}$ and floated on water, with surface tension $\glv=73\mathrm{\,mN\,m^{-1}}$ and  density $\rho=1000\mathrm{~kg\,m^{-3}}$ \footnote{Note that one experiment was performed with a water-glycerol mix with $\glv=68\mathrm{\,mN\,m^{-1}}$ and $\rho=1130\mathrm{~kg\,m^{-3}}$ to test the role of liquid viscosity.}. Steel spheres of radii $0.5\mathrm{~mm}\leq\rs\leq 5.0\mathrm{~mm}$ and density $\rho_s= 7720\mathrm{~kg\,m^{-3}}$ were used as impactors, positioned above the centre of the sheet in a guiding tube (to ensure that the impact occurred vertically), and released using an electromagnet. The impact and resulting sheet deformation were imaged from below using a high-speed camera.  The impact speed $0.6\mathrm{~m\,s^{-1}}\leq V\leq2\mathrm{~m\,s^{-1}}$ was measured by imaging the fall of the sphere from the side. Here, we focus on the deformation of the sheet that occurs at sufficiently early times that the velocity of the sphere is not noticeably affected by impact~\footnote{See Materials and Methods section for further details of the experimental procedure and the SI Appendix (published version) for details of image processing techniques and information about the Supplementary Movies.}.

The images of Fig.~\ref{fig:colour} show two key dynamic features of  impact. 
Firstly, the gross vertical deflection of the sheet (i.e.~the shape on which the wrinkles are superimposed) takes the form of a radially propagating transverse wave.  This wave is analogous to the ripple observed when a stone is dropped into a pond.
Secondly, wrinkles form in the flat region ahead of the transverse wave, and their wavelength gradually increases --- the wrinkles coarsen --- see Figs \ref{fig:colour} and \ref{fig:schematic}b. This is qualitatively different from static indentation experiments \cite{Paulsen2016}, which show that wrinkles maintain a constant wavelength with increasing sheet deformation in the flat portion of such a sheet.   
We note two additional departures from static behaviour: in static indentation experiments, wrinkles are initially confined to a narrow annulus \cite{Vella2015,Ripp2018}, and reach the edge at indentation depths $\sim\, 300\mathrm{~\mu m}$~\cite{Vella2018}, while folds appear at depths $\sim\, 600\mathrm{~\mu m}-2\mathrm{~m m}$~\cite{Holmes2010,Paulsen2016}.  Both of these features appear at impact depths smaller by an order of magnitude ($Vt \sim 100 \mathrm{~\mu m}$) in our dynamic experiments (see Fig.~\ref{fig:schematic}b). 
In this paper we will focus on explaining and quantifying wrinkle coarsening in the flat portion of the sheet during impact; to do so we must first address the transverse wave that drives compression and hence wrinkling in the sheet.   

\begin{figure*}[ht]
\begin{center}
\includegraphics[width=0.97\linewidth]{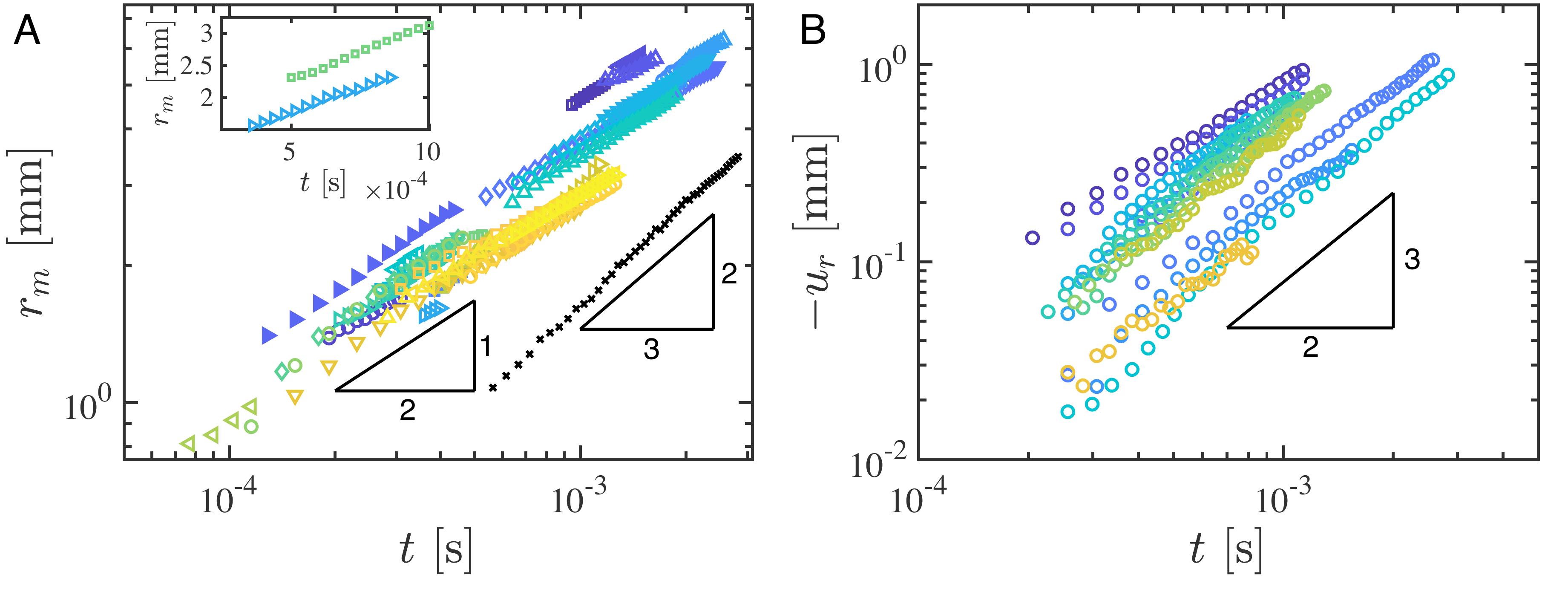}
{\caption{The propagation of the transverse wave and subsequent retraction of the sheet edge. (a) Measurements of the radial position of a capillary wave on a bare interface ($\times$) gives $r_{m} \propto t^{2/3}$, different from the behaviour with that for an ultra-thin PS sheet (colored points). Inset:  wave propagation depends on $\Rf$ (= 5.51\,mm for \textcolor{marker12}{$\rhd$} and 13.30\,mm for \textcolor{marker22}{$\square$}).  (b) The sheet's edge moves radially by an amount $u_r(\Rf) \propto t^{3/2}$, corresponding to  negligible radial stretching. (Experimental parameters for the data in panels a and b are given in tables S1 and S2 of the SI Appendix, respectively.) }
\label{fig:transversewave} }
\end{center}
\end{figure*}

Quantitative results for the propagation of the transverse wave are shown in Fig.~\ref{fig:transversewave}a. While the ripples created by dropping a stone into a pond are known to progress according to the inertia--capillary scaling $r_m\propto t^{2/3}$ \cite{Keller1983,Vandenberghe2016} (see black crosses in Fig.~\ref{fig:transversewave}a), we see that in the presence of an ultra-thin elastic sheet,  $r_m \propto t^{1/2}$ instead. This scaling is  reminiscent of the impact of a sphere into a liquid in the absence of surface tension, for which the contact point between the liquid and solid $r_c = \sqrt{3}(R_s V t)^{1/2}$ \cite{Wagner1932,Philippi2016}. However, the behaviour observed here is distinct from this impact phenomenology since, for example, the experimentally-measured prefactor in the scaling is dependent on the sheet radius $\Rf$ (see inset of Fig.~\ref{fig:transversewave}a). To explain this scaling, we consider the behaviour at very early times, when vertical deflections are small and the effect of gravity on the fluid may be neglected.  Since the fluid is initially at rest and the impact is fast (the Reynolds number $\operatorname{Re}=\rho VR_s/\mu \sim10^3$), we assume that the fluid velocity $\mathbf{u}=\nabla\varphi$ for some  velocity potential $\varphi$. The speed of sound in both the solid and liquid is $O(10^3)~\mathrm{m}\,\mathrm{s^{-1}}$, so that the time scale for sound waves to traverse the  diameter of the sheet is $\tsound=\Rf/c\sim10\mathrm{~\mu s}$. We consider time scales $t\gg\tsound$ so that the fluid is incompressible and 
\begin{equation}
\nabla^2\varphi = 0.
\label{eq:laplace}
\end{equation} 
At the deformed interface $z = w(r,t)$, we impose a kinematic condition
\begin{equation}
\frac{\partial \varphi}{\partial z} = \frac{\partial w}{\partial t} + \frac{\partial \varphi}{\partial r}\frac{\partial w}{\partial r},
\label{eq:kinematic}
\end{equation} 
together with a dynamic boundary condition relating the stresses $\sigma_{rr}$ and $\sigma_{\theta\theta}$ in the sheet to the pressure on the interface $p[r,w(r,t),t]$ via the membrane equation
\begin{equation}
p = -\sigma_{rr}\frac{\partial^2 w}{\partial r^2} - \sigma_{\theta\theta}\frac{1}{r}\frac{\partial w}{\partial r}.
\label{eq:dynamic}
\end{equation}   
The pressure $p$  is, in turn, related to the velocity potential $\varphi$ through Bernoulli's equation
\begin{equation}\label{eq:bernoullia}
\rho \left( \frac{\partial \varphi}{\partial t} + \frac{1}{2} |\nabla\varphi|^2 \right) + p = 0.
\end{equation} 
In writing~\eqref{eq:dynamic} we have neglected both the bending stiffness and the inertia of the sheet.  The PS sheets used in our experiments are very thin and therefore highly bendable~\cite{Davidovitch2011,Vella2015,Paulsen2016}; bending stiffness is negligible over the lengthscale of the transverse wave, although we shall see below that it is important in selecting the wrinkle wavelength. 

Highly bendable sheets cannot sustain compressive stresses.
Instead, wrinkles form very early on, relaxing the compressive hoop stress so that $|\sigma_{\theta\theta}| \ll \sigma_{rr}$.
Our experiments focus on times $t\gg\tsound$, for which in-plane stresses are in equilibrium, and so $\sigma_{rr} \approx \glv \Rf/r$  \cite{Davidovitch2011,Vella2015} (distinguishing our experiments from many previous studies of dynamic buckling \cite{Karagiozova2008,Gladden2005,Vermorel2007,Vermorel2009} and especially ref.~\cite{Vandenberghe2016} where $t\lesssim \tsound$).  
 From a scaling point of view, Laplace's equation~\eqref{eq:laplace} suggests that the vertical length scale over which the (infinite) bath of fluid feels the impact $z_\ast\sim r_m$. The kinematic and dynamic boundary conditions,~\eqref{eq:kinematic} and \eqref{eq:dynamic}--\eqref{eq:bernoullia} respectively, then suggest that $\varphi_\ast \sim r_m V \sim \glv \Rf V t^2/(\rho r_m^3)$. Combining these scalings, we find that
\begin{equation}
\rm \propto \left(\frac{\glv \Rf t^2}{\rho}\right)^{1/4},
\label{eq:wavescaling}
\end{equation}
while the height of the wave scales with $V t$. We emphasize that the predicted scaling $\rm \propto t^{1/2}$ derives from the spatially-varying stress $\sigma_{rr} \propto 1/r$ and requires the sheet to be highly wrinkled with the stresses in equilibrium.  (By contrast, non-equilibrium stresses generated by a concurrent tensile wave instead cause the transverse wave to propagate as $t^{2/3}$ \cite{Vandenberghe2016}.)

\begin{figure*}
\begin{center}
\includegraphics[width=0.97\linewidth]{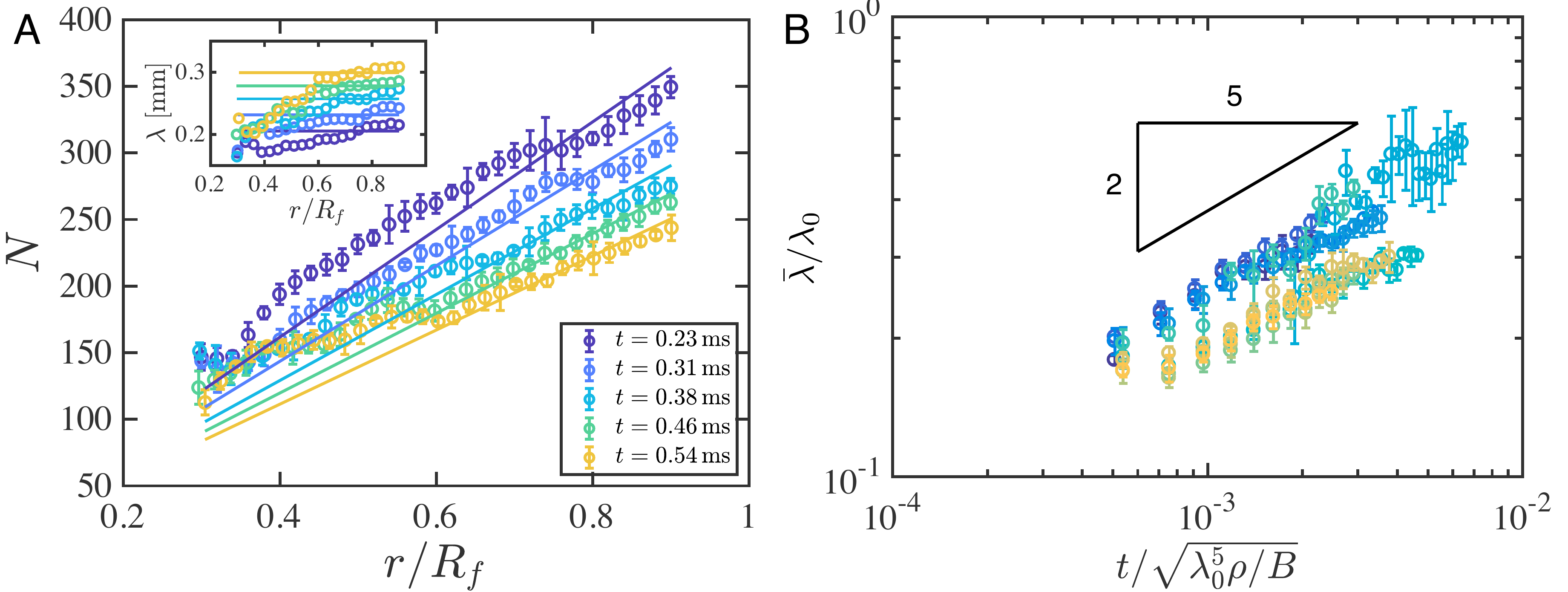}
\caption{The instantaneous and dynamically-evolving wrinkle wavelengths.  (a) The number of wrinkles $N$ increases approximately linearly with radial distance from the point of impact, corresponding to an instantaneous average wavelength that is approximately uniform (inset). Data points show measurements at different times, as indicated in the legend; solid lines indicate linearity, and a fixed value (inset). Here,  $h=450$\,nm, $\Rf=13.3$\,mm, $R_s = 1.25$\,mm and $V=1.11$\,m\,s$^{-1}$. (b) The mean wavelength $\lambdabarR$ of the radial wrinkles in a fixed material circle, $r(t) = 0.8\bigl[\Rf-u_r(\Rf,t)\bigr]$, increases with time. $\lambdabarR$ was measured by counting the number of wrinkles and averaging azimuthally; $\lambdabarR$ is normalized with the static wavelength $\lambda_0$ given by~\eqref{eq:gravity}.    Experimental parameters for panel b are given in table S3 of the SI Appendix. In (b) the scaling of \eqref{eq:wrinklescaling} is highlighted by the triangle.  The errorbars represent the standard deviation of measurements obtained in ten intensity signals from neighbouring pixel strips of an unwrapped image (Fig.~4a) or spatio-temporal (Fig.~4b) plot (see Supplementary Information for further details).  Here, the axes are scaled with the static wavelength $\lambda_0 \approx 1.5$--$1.7 \mathrm{~mm}$ and the corresponding timescale $(\lambda_0^5 \rho/B)^{1/2} \approx 0.7\, \mathrm{s}$.}
\label{fig:wrinkling}
\end{center}
\end{figure*}

While the impacted sheet might be expected to stretch, in fact $\gamma/(Eh)\sim10^{-4}\ll1$ so that the sheets are effectively inextensible. 
This preserves the length of radial lines (i.e.\ the radial strain $\epsilon_{rr}\approx0$), so that the edge retraction follows directly from the vertical deflection, $-u_r(\Rf)\approx\tfrac{1}{2}\int_0^{\Rf}(\partial w/\partial r)^2~\upd r\sim (Vt)^2/r_m$.  Using \eqref{eq:wavescaling}, this yields
\begin{equation}
-u_r(\Rf) \propto V^2 \left(\frac{\rho}{\gamma \Rf}\right)^{1/4} t^{3/2}.
\label{eq:recedingedge}
\end{equation} 
This scaling prediction is confirmed by experimental data (Fig.~\ref{fig:transversewave}b). 
We note that our assumption of inextensibility (and hence the calculation of radial retraction) is only valid because the dominant tension arises from capillarity, in contrast to the dynamic indentation experiments of~\cite{Vandenberghe2016} where impact induces significant stretching of the sheet.  We also note that, although the power laws predicted by Eqs.~(\ref{eq:wavescaling}--\ref{eq:recedingedge}) and illustrated in Fig. 3 appear to be robust, there is some spread in the prefactor; we show elsewhere \cite{prfluids} that this spread in prefactor is largely accounted for by the size of the impactor. 

The inward displacement of the sheet edge is crucial in driving wrinkling: radial retraction of the sheet $u_r(t)$  leads to an azimuthal compression, which is relieved by the growth of wrinkles that accommodate excess material.   Experimentally, we observe that the number of wrinkles in a material circle increases approximately linearly with radial distance from the point of impact (Fig.~\ref{fig:wrinkling}a), so that the azimuthally-averaged wavelength $\lambdabarR(t)$ is approximately uniform in the outer region of the sheet at each instant (inset of Fig.~\ref{fig:wrinkling}a).  This breaks down in the inner region of the sheet where radial tension and curvature from the transverse wave provide additional stiffnesses~\cite{Paulsen2016}.  The dynamic evolution of $\lambdabarR(t)$ in the flat region of the sheet is shown in Fig.~\ref{fig:wrinkling}b, and demonstrates that the average wavelength increases (or, equivalently, that the number of wrinkles decreases) with time and is smaller than the corresponding static wavelength governed by the fluid's hydrostatic pressure~\eqref{eq:gravity}.

To understand wrinkle formation and coarsening, we consider a material circle in the flat portion of the sheet, ahead of the transverse wavefront.  We develop a simplified model by considering wrinkling in this material circle, driven by the compressive stress arising from radial retraction, and moderated by the bending stiffness $B = E\h^3/[12(1-\nu^2)]$ of the sheet and the inertia of the underlying fluid.  We focus on large radial positions only, so that the curvature of the sheet (associated with the transverse wave) may be neglected, and $r \gg \lambda$.  We emphasize that the previously neglected bending stiffness of the sheet and small residual compressive hoop stress must both be accounted for over the short lengthscale associated with wrinkling.  We therefore model a freely floating 1D sheet, which is subject to a compressive force $P(t)$ that mimics the compressive hoop stress $\sigma_{\theta\theta}(t)$, and evolves as a result of an imposed compressive displacement of its ends $\Delta(t)$.
The wrinkle coarsening is the result of the sheet being unable to move the liquid instantaneously, but, unlike previous studies \cite{Kodio2017,Chopin2017}, here the fluid flow is inertial, not viscous. 

We consider a sheet that lies along the $x$-axis and model its out-of-plane displacement, $w(x,t)$, using the beam equation subject to a linearized hydrodynamic pressure from \eqref{eq:bernoullia}. A balance between the bending stress, compressive stress and flow induced by wrinkling gives 
\begin{equation}
-\rho\frac{\partial\varphi}{\partial t} = B\frac{\partial^4 w}{\partial x^4} + P(t)\frac{\partial^2 w}{\partial x^2}.
\label{eq:beam}
\end{equation}
The compressive force $P(t)$ is not known \emph{a priori}, and is determined by imposing a confinement
\begin{equation} \label{eq:geometric}
\int_{-\pi r}^{\pi r} \left(  \frac{\partial w}{\partial x} \right)^2 ~\mathrm{d} x = \Delta(t),
\end{equation} which expresses that the sheet wrinkles without changing its length.   
From a scaling point of view, the kinematic boundary condition gives $\partial\varphi/\partial z \sim \partial w/\partial t$. A scaling analysis of \eqref{eq:beam} then gives
\begin{equation}
\lambda \propto \left(\frac{B}{\rho}\right)^{1/5} t^{2/5},
\label{eq:wrinklescaling}
\end{equation} 
so that the wavelength is selected simply by a balance between bending stiffness and a `dynamic substrate stiffness' associated with fluid inertia. The scaling  \eqref{eq:wrinklescaling} gives a reasonable account of our experimental data (see Fig.~\ref{fig:wrinkling}b). We note that the radial position $r$ and confinement $\Delta(t)$ affect the wrinkle amplitude through~\eqref{eq:geometric} but not the scaling argument~\eqref{eq:wrinklescaling}, which is derived from~\eqref{eq:beam} only.  Furthermore, the length constraint~\eqref{eq:geometric} prevents the existence of an exact similarity solution of the form~\eqref{eq:wrinklescaling}; numerical solutions presented in ref.~\cite{prfluids} demonstrate that the scaling of \eqref{eq:wrinklescaling} may have a logarithmic-type correction.


We have studied the dynamic wrinkling of an ultrathin elastic sheet floating on a liquid interface, highlighting several key features of this motion.  The motion $r_m\propto t^{1/2}$ of the transverse wave illustrates the large change in the state of stress caused by highly developed wrinkling, while the evolution of the wrinkle pattern is very different to that observed statically \cite{Vella2015,Paulsen2016}:  fluid inertia slows out-of-plane deformation of the sheet, selecting a dynamically evolving, but almost spatially uniform wrinkle wavelength. 
Moreover, the effect of the fluid inertia can be understood as the result of a dynamic substrate stiffness, $\Kdyn$, which supplements the substrate-, tension- and curvature-induced stiffnesses that have been introduced for static wrinkle patterns \cite{Cerda2003,Davidovitch2011,Paulsen2016}. In particular, \eqref{eq:wrinklescaling} can be rewritten as $\lambda \propto (B/\Kdyn)^{1/4}$, with $\Kdyn \sim \rho\times\lambda/t^2$ arising from the hydrodynamic pressure associated with accelerating liquid to accommodate increasing $\lambda$: by addressing a dynamic scenario we effectively introduce a new source of stiffness that competes with the bending stiffness of the sheet to control the wrinkle wavelength.  We note that at sufficiently late times, this additional hydrodynamic pressure will decrease below the typical hydrostatic pressure, ultimately leading to quasi-static wrinkles governed by the usual static stiffness $K_0=\rho g$. 

A surprising feature of our main results, \eqref{eq:wavescaling} and \eqref{eq:wrinklescaling}, is that they do not have explicit dependencies on the impact velocity $V$. This is because $V$ affects the magnitude of the vertical displacement of the sheet, but not  how fast this disturbance propagates: that speed is instead set by the intrinsic properties of the sheet and liquid. Note, however, that the radial displacement at the edge of the film, \eqref{eq:recedingedge}, \emph{does} depend on $V$, since it is induced by the vertical motion. Nevertheless, the wrinkle wavelength $\lambda$ is independent of $V$, since the amount of compression to be accommodated merely changes the amplitude of wrinkles.

We expect the dynamic picture we have presented here to hold while the transverse wave is travelling across the sheet, i.e.~while $\rm\ll\Rf$, which in turn requires $t\ll(\rho \Rf^3/\gamma)^{1/2}$. At the same time, we also need to ensure that the dynamic wrinkle wavelength $\lambda\ll \lambda_0=(B/\rho g)^{1/4}$, which amounts to the requirement that $t\ll(\lambda_0/g)^{1/2}$ (note that the data in Fig.~\ref{fig:wrinkling}b all readily satisfy this constraint). Both of these results are independent of the velocity of the impactor, $V$. The key point that determines whether an impact is inertial in the sense considered here is whether the vertical distance travelled during this early time is large compared to the critical vertical displacement at which wrinkling occurs in the static problem, $\delta_c \sim \gamma/(Eh\rho g)^{1/2}$ \cite[see refs.~][for example]{Vella2015,Vella2018}. This comparison gives us that the wrinkling of the sheet is governed by the inertia of the fluid if 
\beq
V\gg V_c=\left(\frac{\gamma^2}{Eh\cdot\rho}\right)^{1/2}\max\left[\left(\frac{\gamma}{\Rf^3 \rho g}\right)^{1/2},\lambda_0^{-1/2}\right];
\eeq for the parameters typical of the floating PS sheets considered in our experiments, $V_c=O(1\mathrm{~mm}\,{\mathrm{s}^{-1}}{)}$ so that the impacts we consider, with $V=O(1\mathrm{~m}\,{\mathrm{s}^{-1}})$, all lie well within this regime.

Our experiments suggest that a dynamic substrate stiffness may provide a means of breaking away from the single, static wavelength that is selected by material properties alone, allowing the wavelength to be altered without resorting to non-uniform substrates or coatings \cite{Schleifer2019}.  This is a new route for tunable wrinkle formation that may prove to be a useful fabrication technique in a range of engineering applications that require regular, patterned topographies \cite{Schweikart2009,Yang2010}. In particular, we have shown that this wavelength change can occur very quickly, with a doubling of the mean wavelength in around $1\mathrm{~ms}$  (see Figs.~\ref{fig:schematic}b and ~\ref{fig:wrinkling}b).

The rapid coarsening of wrinkle wavelength that we have presented occurs with wavelengths on the order of $100\mathrm{~\mu m}$, making it readily observable.  However, the underlying mechanism  is scale-independent, provided that the Reynolds number of the fluid flow remains sufficiently large that inertia dominates viscosity. This mechanism might therefore be suitable for reproduction at similar scales but with still thinner sheets to obtain wrinkles at smaller length scales. A key objective would be to produce wrinkle wavelengths small enough for use in applications with visible light. For example, wrinkle wavelengths around $\lambda\approx 4.7 \mu \mathrm{m}$ were used in ref.~\cite{Li2013} to focus light; our model suggests that extremely thin ($h \sim 7 \mathrm{n}\mathrm{m}$)  polystyrene sheets~\cite{Chang2018} would  generate wrinkles with similar wavelengths over a fraction of a millisecond. Many optical applications, including photonic materials \cite{Fudouzi2003} and Bragg gratings \cite{Shen2014}, require periodic structures with period comparable to the wavelength of visible light. If our impact protocol were implemented with monolayer graphene floating on water (with bending rigidity $B\sim10^{-19}\mathrm{~J}$ \cite{Lu2009}), our theory predicts that the wrinkle wavelength would double from $\lambda\approx400\mathrm{~nm}$ to $\lambda\approx800\mathrm{~nm}$ (thereby transitioning from the wavelength of blue light to beyond that of red light) on a time scale of $\sim 2\mathrm{~\mu s}$.

Applications of dynamic wrinkling would benefit from other means of generating the rapid azimuthal compression required for wrinkling. While high-speed indentation with a linear actuator could replace impact, the negative thermal expansion coefficient of graphene \cite{Yoon2011} could also be exploited; indeed, contraction caused by ultrafast optical excitation can create a compressive strain of $6\times10^{-4}$ within $100\mathrm{~ps}$~\cite{Hu2016}. This compressive strain would be sufficient to overcome the isotropic tension caused by surface tension, $\gamma/(Eh)\approx 2\times10^{-4}$, and could therefore be used to induce dynamic wrinkling.  The practicalities of achieving such rapid wrinkling in ultra-thin sheets deserve further investigation.



\small
\subsection*{Materials}
Polystyrene (PS) sheets were created by dissolving PS powder (Goodfellow) in toluene (anhydrous, 99.8\%, Sigma-Aldrich) and  spin coating the solution onto glass slides \cite{Huang2007} at 1000\,rpm for 60\,s (spin coater Polos SPIN150i). Films of different thicknesses, in the range $150\mathrm{~nm}\leq \h\leq 530\mathrm{~nm}$, were created by varying the concentration of the solution (from $2-4.5$\,g PS per $100\mathrm{~mL}$ toluene), and were measured using a thin film analyser (F20, Film Metrics). 
 The resulting sheets were cut into circular disks using a diamond-tipped cutting tool and floated on water, with reported surface tension  $\glv \approx 73\mathrm{\,mN\,m^{-1}}$,  viscosity $\mu\approx0.89$\,mPa\,s and  density $\rho\approx1000\mathrm{~kg\,m^{-3}}$. The resulting sheet radius was measured from images and was varied in the range $5.5\mathrm{~mm}\leq\Rf\leq22.7\mathrm{~mm}$.

\subsection*{Methods}
Impact experiments were performed by dropping steel spheres (Simply Bearings), radii $0.5\mathrm{~mm}\leq\rs\leq 5.0\mathrm{~mm}$ and density $\rho_s= 7720\mathrm{~kg\,m^{-3}}$ onto floating PS films. A schematic of the experimental setup is shown in Fig.~\ref{fig:schematic}, with a video provided as Movie S1. Spheres were released using a custom-built electromagnet, positioned so that spheres impacted the centre of the sheet. A guiding tube was used to ensure that the impact occurred vertically. The impact was imaged from below using a high-speed video camera (Miro 310, Phantom), typically at a frame rate of $39,024\mathrm{~Hz}$ with $256 \times 256$ pixels and with a spatial resolution of $0.02$\,mm per pixel. The impact speed  $0.6\mathrm{~m\,s^{-1}}\lesssim V\leq2\mathrm{~m\,s^{-1}}$ was measured by imaging the fall of the sphere, at $1000\mathrm{~Hz}$, using a second camera (FinePix HS10, Fujifilm).

We performed one impact experiment in which the underlying fluid had a higher viscosity than water, to assess the influence of liquid viscosity.  The liquid was a glycerol-water mixture (55:45\% by volume), with dynamic viscosity $\mu = 13$\,mPa\,s,  surface tension coefficient $\gamma_{lv} = 68$\,mN\,$\mathrm{m}^{-1}$ and density $\rho = 1130$\,kg\,$\mathrm{m}^{-3}$.  The transverse wavefront data for this experiment are illustrated in Fig.~3a of the main text, highlighted with filled-in markers. The results are consistent both with the other experiments and with the theoretical prediction, confirming that viscosity does not play a significant role, validating our neglect of fluid viscosity.

The  measured values of the instantaneous wavelength were distributed around some mean value, rather than being a single well-defined wavelength (SI Appendix); the average wavelength reported herein was determined by counting visible wrinkles in images and averaging  this number over the length of a particular material circle. We investigate the distribution of wavelengths in more detail in a companion paper~\cite{prfluids}.

The results presented  correspond to approximately constant impact velocities, with data only shown up to the time at which the predicted contact radius becomes comparable to the sphere radius~\cite{prfluids}. The only exception to this is in the inset to Fig.~3a, where we have displayed data over a longer time interval to clearly illustrate the dependence on $R_f$. 

\begin{acknowledgments}
The research leading to these results has received funding from  the European Research Council under the European Union's Horizon 2020 Programme / ERC Grant Agreement no.~637334 (DV), the Royal Society (AAC-P) and the OUP John Fell Fund. 
\end{acknowledgments}





\end{document}